\documentclass[aps,twocolumn,groupedaddress]{revtex4}
\usepackage{graphicx,color}
\usepackage{amsmath}
\usepackage{longtable}
\begin{document}

\title{Particle flows in a dc discharge in laboratory and microgravity conditions}

\author{S. Khrapak,$^{1,2}$ M. H. Thoma,$^1$ M. Chaudhuri,$^1$  G. E. Morfill,$^1$ A. V. Zobnin,$^2$ A. D. Usachev,$^2$ O. F. Petrov,$^2$ and V. E. Fortov$^2$}
\affiliation{$^1$Max-Planck-Institut f\"ur extraterrestrische Physik, D-85741 Garching, Germany\\ $^2$Joint Institute for High Temperatures, 125412 Moscow, Russia
}
\date{\today}

\begin{abstract}
We describe a series of experiments on dust particles flows in a positive column of a horizontal dc discharge operating in laboratory and microgravity conditions. The main observation is that the particle flow velocities in laboratory experiments are systematically higher than in microgravity experiments, for otherwise identical discharge conditions. The paper provides an explanation for this interesting and unexpected observation. The explanation is based on a physical model, which properly takes into account main plasma-particle interaction mechanisms relevant to the described experimental study. Comparison of experimentally measured particle velocities and those calculated using the proposed model demonstrates reasonable agreement, both in laboratory and microgravity conditions, in the entire range of discharge parameters investigated.

\end{abstract}

\maketitle

\section{Introduction}\label{intro}

Complex (dusty) plasmas constitute an interdisciplinary research field with relations to space and astrophysical topics~\cite{Goertz}, industrial plasma applications~\cite{Merlino1}, plasma fusion oriented research~\cite{Krasheninnikov}, physics of strongly coupling phenomena~\cite{Bonitz}, and soft condensed matter~\cite{ChaudhuriSM,Book1}. The problem of plasma-particle interactions is central to this  field, because these interactions govern practically all phenomena that can be observed and investigated~\cite{FortovUFN,FortovPR,Book2}. Of particular importance are interactions that affect particle charging and screening~\cite{Book2,KhrapakCPP}, momentum transfer between different complex plasma components~\cite{MT}, external and internal forces acting on the particles~\cite{Book2}.

The focus of this paper is on a situation, where different complex plasma components are allowed to drift relative to each other, in an electric field of a positive column of a direct current (dc) discharge. Particles are injected into a horizontally mounted dc discharge tube and are transported along the tube by various forces, most important of which are the electrical, ion drag, and neutral drag forces. The particle velocities can be relatively easily measured in a wide range of discharge parameters and they provide important information on various basic complex plasma properties, in particular particle charges and forces they are acted upon. A series of experiments that we describe here is performed using the same experimental setup in laboratory and under microgravity conditions (during the parabolic flights). The main observation is that {\it particle velocities are systematically higher in laboratory as compared to microgravity conditions} at otherwise identical discharge conditions. This finding is naturally a puzzling one, since the plasma parameters (and hence plasma-particle interaction mechanisms) are not expected to depend on the presence or absence of gravity. The main purpose of the present paper is to provide a convincing explanation for this unexpected observation.

We analyze in detail the specifics of particle flows in a positive column of a horizontal dc discharge in ground-based (laboratory) and microgravity conditions. In doing so we put forward a physical model, which is believed to correctly describe main plasma-particle interactions relevant to the present experiment. The issue of particle charging in a flowing collisional plasma is the main constituent of the model. We show that in laboratory, the force of gravity shifts particles downwards from the vicinity of the tube axis. As a result, parameters of the plasmas {\it surrounding} the particles in laboratory and microgravity conditions are somewhat different. In particular, particles in laboratory experiments are located in a region with higher ion drift velocity. Higher ion drift velocities result in higher particle charges and higher electrical forces acting on the particles. This is the main qualitative explanation behind the difference in particle velocities measured in laboratory and microgravity conditions. We finally demonstrate reasonable quantitative agreement between experimental measurements and the analytical theory proposed in this paper.

The paper is organized as follows. In Section~\ref{experiment} we describe the experimental apparatus, experimental procedures, and main observations. Section~\ref{Model} introduces the model of particle charging relevant to the present experimental conditions. In Section~\ref{Forces} we summarize main forces acting on the particles and in Section~\ref{Force-balance} we consider the force balance that determines the particle motion, separately for laboratory and microgravity conditions. Results from this analysis are discussed in Section~\ref{results}. This is followed by conclusion in Section~\ref{conclusion}.

\section{Experiment}\label{experiment}

\begin{figure}
\centering
\includegraphics[width=7.4cm]{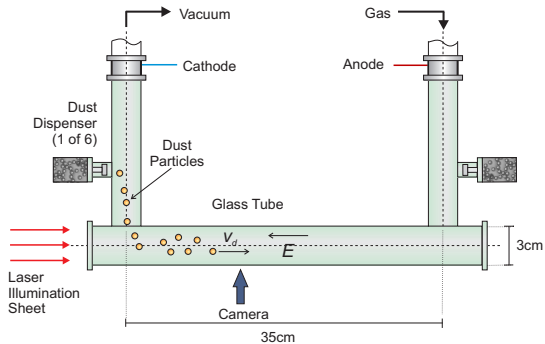}
\caption{Sketch of the PK-4 facility in the horizontal configuration.}\label{sketch}
\end{figure}

The experiments are carried out in the Plasma Kristall-4 (PK-4) facility, scheduled for operation onboard the International Space Station (ISS) after 2014 \cite{PK4_a,PK4_b}.  It represents a dc discharge generated in a U-shaped glass tube of 3 cm in diameter. The main part of the tube containing the positive column of the discharge is 35 cm long. In the described experiments the horizontal configuration of the tube is employed (see Fig.~\ref{sketch}). Complex plasma is formed by injecting $\mu$m-size grains into the discharge. The particles can be observed and recorded by video cameras at a required position inside the tube. Normally, individual particle resolution is possible. The sketch of the PK-4 experimental setup is shown in Fig.~\ref{sketch}. For a comprehensive description of this project see e.g. Refs.~\cite{PK4_a,PK4_b}.

The PK-4 facility is particularly suited to study different kinds of flow-related phenomena in complex plasmas. Previous experiments in laboratory and microgravity (parabolic flights) conditions have already provided a wealth of useful information, in particular, regarding particle charging~\cite{RatynskaiaPRL,KhrapakPRE,KhrapakEPL}, the ion drag force~\cite{YaroshenkoPoP}, particle-particle interactions~\cite{ZobninPRL,IvlevPRL2011}, linear and non-linear waves~\cite{RatynskaiaIEEE,ZhdanovEPL}, structural properties of the particle component~\cite{Slobo1,Slobo2}, and other important phenomena~\cite{OverviewPK4}. Experiments reported here correspond to a series of basic studies, aiming at checking the reliability of the PK-4 setup under microgravity conditions

\begin{figure*}
\centering
\includegraphics[width=16cm]{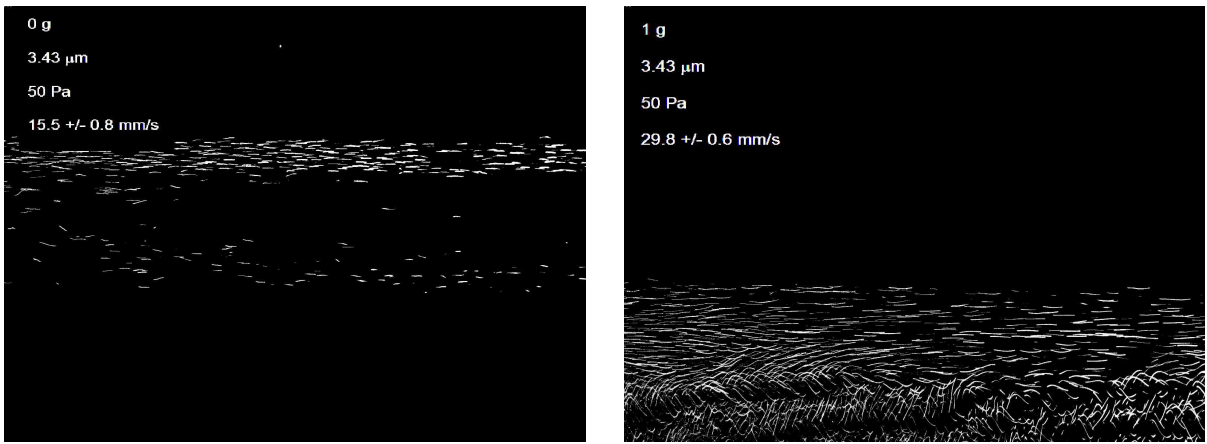}
\caption{Example of the particle flow in microgravity (left panel) and laboratory (right panel) experiments. Both snapshots correspond to the particle size $2a=3.43$ $\mu$m and neutral gas pressure $p=50$ Pa. Particles are drifting to the left. The particle flow velocity in laboratopry conditions is almost a factor of two higher (note longer track lengths). Note also that in laboratory conditions the particle cloud is considerably shifted (downwards) from the tube axis, which corresponds to the centers of the images. Waves visible in the lower part of the right panel are related to an increase in the total electric field (radial field is required to levitate the particles in laboratory conditions). These waves are excited by sufficiently fast ion flows and the corresponding ion-dust streaming instability~\cite{Rosenberg,FortovPoP_2000,Joyce,Merlino2}. The field of view is $22.1\times 16.6$ mm$^2$ ($1600 \times 1200$ pixels$^2$). The exposure time is 28 ms.}\label{Flows}
\end{figure*}

The experiments have been performed in laboratory conditions and under microgravity,  during the parabolic flights campaign in the fall 2012. For these experiments neon gas in the pressure range between $\sim 30$ and $\sim 90$ Pa is used, the discharge current is fixed at $1$ mA. Spherical melamine-formaldehyde particles (mass density is $\rho\simeq 1.51\,$g/cm${^3}$) are injected into the discharge tube. Being injected into the plasma, the particles become charged negatively and are transported through the positive column of the discharge by the combination of various forces. The dominant forces in the horizontal direction are the electrical, ion drag and neutral drag forces~\cite{RatynskaiaPRL,KhrapakPRE,KhrapakEPL}. The particles are illuminated by a laser sheet and their motion is recorded by a video-camera, situated near the center of the tube. The video-camera has a field of view of $22.1\times 16.6$ mm$^2$ and operates at a frame rate $35$ fps.  Figure \ref{Flows} shows typical snapshots of drifting particle clouds in microgravity (left panel) and laboratory conditions (right panel).


The velocities of the particles are estimated by measuring their track lengths in single images (snapshots). Each experimental point corresponds to the average value of track lengths selected randomly in different images for a particular set of experimental parameters (neutral gas pressure, particle size, particle dispenser). The procedure is essentially the same as used previously~\cite{RatynskaiaPRL,KhrapakPRE}.

The main observations are as follows: (i) For any particular particle size the particle drift velocities decrease monotonically with increasing pressure; (ii) For a given particle size and pressure, the drift velocities in laboratory experiments are systematically higher than in microgravity conditions; (ii) The difference between drift velocities in laboratory and microgravity conditions increases with increasing the particle size and with lowering neutral gas pressure.

To proceed with quantitative explanation of these observations,  the knowledge of main plasma parameters is required. The plasma density, electron temperature, and axial electric field were measured previously in the absence of particles~\cite{PK4M}  using a single Langmuir probe of $\simeq 4$ mm length and $\simeq 30$ $\mu$m in diameter, insulated by a glass holder~\cite{Zobnin_probes}. The linear fits to the experimental results applicable in the pressure range investigated were given in Ref.~\cite{KhrapakPRE}. Since a relatively small number of particles is injected in each experimental run and the particle cloud extent in the radial direction is considerably smaller than the tube radius (see Fig. \ref{Flows}), we neglect all kinds of collective effects as well as the effect of the particles on plasma parameters. The analysis is thus performed in the individual particle approximation.

\section{Model of particle charging}\label{Model}

The purpose of this Section is to present a simple heuristic model of particle charging in collisional plasmas with flowing ions, which can be used to explain the observations summarized above. Two important effects should be accounted for in such a model: Ion-neutral collisions and relative drift between the ions and the particle. Although, the effect of ion-neutral collisions on the particle charging has received considerable attention~\cite{Zobnin,Lampe,Hutchinson1,Dyachkov,Zobnin1,Interpol,Semenov} (for a review see Ref.~\cite{KhrapakCPP}), most of the studies have concentrated on isotropic conditions. To the best of our knowledge there is very few papers, which reports results for ion collection by a sphere in a {\it drifting} collisional plasma~\cite{HutchinsonID,Haakonsen}. Numerical results from Ref.~\cite{Haakonsen} are more comprehensive from the point of view of comparison, and thus they will be used to check the reliability of the approximation we put forward here.

The reduced particle surface potential $z= e|\phi_{\rm s}|/T_e$ (and thus the particle charge $Q\simeq \phi_{\rm s}a$) is set by the balance between the electron and ion fluxes absorbed on the particle surface (here $\phi_{\rm s}$ is the particle surface potential, $a$ is the particle radius, $T_e$ is the electron temperature, and $e$ is the elementary charge). In order to simplify the comparison with Ref.~\cite{Haakonsen} we use the following normalization for the fluxes: The fluxes are in units of $4\pi n_i a^2 C_{\rm s}$, where $n_i$ is the ion density and  $C_{\rm s}=\sqrt{T_e/m_i}$ is the ion sound speed. In the pressure range investigated, electron-neutral collisions are not important and the conventional OML expression for the electron flux~\cite{OML} is applicable,
\begin{equation}\label{Je}
j_e=\sqrt{m_i/2\pi m_e}\exp{(-z)}\,.
\end{equation}

In order to derive an approximation for the ion flux we consider separately two regimes. In the first -- {\it weakly collisional (WC) regime} -- a reasonable approximation for the ion flux is a linear superposition of the collisionless (OML) and collisional contributions~\cite{Lampe,Hutchinson1}. The OML expression for the shifted Maxwellian distribution reads in the present notation as
\begin{equation}
\begin{array} {lcl}
j_i^{\rm OML} & = & \frac{1}{4\xi\sqrt{2\pi\tau}}\left[\sqrt{\pi}\left(1+2\xi^2+2z\tau\right){\rm erf}(\xi) + \right. \\ \\ & + & \left. 2\xi\exp(-\xi^2)\right] ,
\end{array}
\end{equation}
where $\xi=M\sqrt{\tau/2}$ is the measure of the ion drift velocity expressed in terms of the Mach number, $M=u/C_{\rm s}$, and $\tau=T_e/T_i$ is the electron-to-ion temperature ratio. The collisional contribution to the ion flux in the {\it isotropic} situation can be roughly estimated (in dimensional units) as $\Delta J_i^{\rm coll}\simeq (4\pi/3)R_0^3n_i\nu$, where $R_0$ denotes the radius of a sphere inside which the ion-particle interaction is sufficiently strong~\cite{Lampe,KhrapakPoP2012} and $\nu$ is the effective ion-neutral collision frequency. For the Debye-H\"{u}ckel (Yukawa) potential a simple formula of the type $R_0\simeq \lambda\ln(1+R_{\rm C}/\lambda)$, where $\lambda$ is the plasma screening length and $R_{\rm C}=|Q|e/T_i$ is the (ion) Coulomb radius, would describe adequately the respective limits of weak and strong ion-particle coupling and provide a smooth transition between them~\cite{KhrapakPRE2012}. The remaining step is to estimate how the presence of the ion flow modifies this collisional contribution.

In the spirit of Ref.~\cite{KhrapakPoP2005} we define the effective Coulomb radius and plasma screening length in flowing plasma as follows. The Coulomb radius is $R_{\rm C}\simeq \frac{|Q|e}{T_i(1+\xi^2)}$ so that it reduces to the standard definition in isotropic situation ($\xi=0$) and decreases as the ion kinetic energy (associated with the direct ion flow) increases. The effective plasma screening behaves as $\lambda\simeq \lambda_{{\rm D}e}/\sqrt{1+f(\xi)\tau}$, where $f(\xi)=(1+2\xi^2)^{-1}$ is an adjusting function and $\lambda_{{\rm D}e}=\sqrt{T_e/4\pi e^2 n_e}$ is the electron Debye radius. This ensures that $\lambda$ tends to $\lambda_{{\rm D}i}=\sqrt{T_i/4\pi e^2 n_i}$ in the isotropic regime, approaches $\lambda_{{\rm D}e}$ for highly suprathermal flows, and provides smooth transition between these limits~\cite{KhrapakPoP2005,Ludwig}.  With the use of these approximations, the collisional contribution to the ion flux can be written in the present (dimensionless) notation as
\begin{equation}
\Delta j_i^{\rm coll}\simeq \frac{\omega_c L_e^3}{3[1+f(\xi)\tau]^{3/2}}\ln^3\left[1+\frac{z\tau\sqrt{1+f(\xi)\tau}}{L_e(1+\xi^2)}\right],
\end{equation}
where $\omega_c=\nu a/C_{\rm s}$ is the reduced ion-neutral collision frequency (ion collisionality index) and $L_e=\lambda_{{\rm D}e}/a$. The total flux that the particle collects in the weakly collisional regime is simply $j_{\rm WC}\simeq j_i^{\rm OML}+\Delta j_i^{\rm coll}$.

In the {\it highly collisional (HC) regime} we make use of the problem of collisional ion flow around a Coulomb sphere, which allows for an analytical solution~\cite{HutchinsonComment}. In the present notation the result can be written as:
\begin{equation}
j_{\rm HC} = \begin{cases} z/\omega_c, & M\leq z/\omega_c \\ \frac{M}{4}\left(1+z/M\omega_c\right)^2, & M>z/\omega_c \end{cases}
\end{equation}

Finally, the interpolation formula, which is applicable in the entire range of ion collisionality is
\begin{equation}\label{Jeff}
j_i^{\rm eff}=\left(j_{\rm WC}^{-\gamma}+j_{\rm HC}^{-\gamma}\right)^{-1/\gamma}
\end{equation}
where $\gamma$ is generally an adjustable parameter~\cite{Hutchinson1}. The simplest choice $\gamma =1$ has been shown to yield reasonable results in many cases~\cite{Interpol} and we adopt it here, although it should be realized that treating $\gamma$ as a free parameter can naturally provide better accuracy~\cite{Hutchinson1,KhrapakEPL}.

\begin{figure}
\centering
\includegraphics[width= 8 cm]{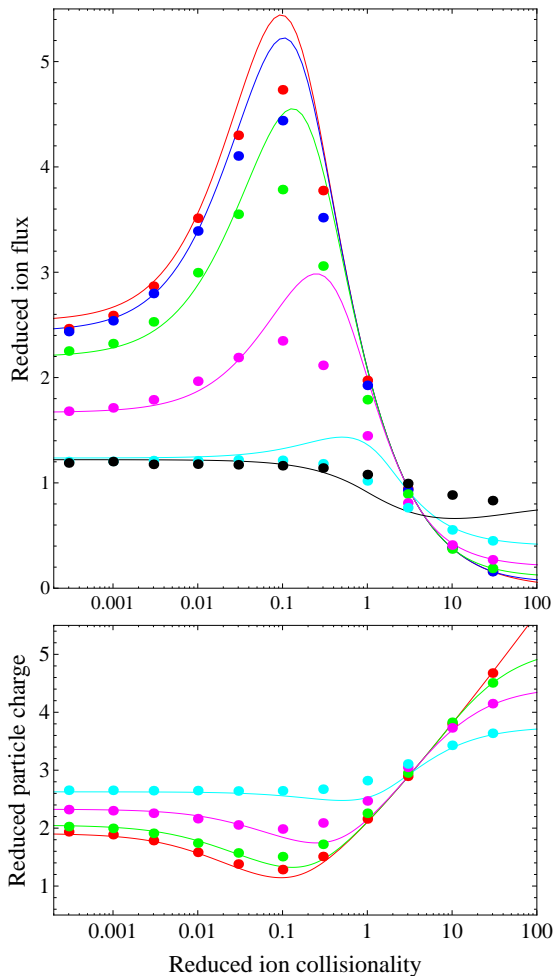}
\caption{(Color) Normalized ion flux to the particle surface $j_i$ (top panel) and reduced particle charge $z$ (bottom panel) vs. the reduced ion-neutral collision frequency $\omega_c$ for several ion drift velocities. Symbols correspond to the numerical results from Ref.~\cite{Haakonsen}; solid curves are computed using the present approximation [Eq.~\ref{Jeff}]. At the left border of the top panel the symbols and curves from top to bottom correspond to $M=0.0$ (red), $M=0.2$ (blue), $M=0.4$ (green), $M=0.8$ (magenta), $M=1.6$ (cyan), and $M=3.2$ (black). Similarly, at the left border of the bottom panel the curves and symbols from top to bottom are for $M=1.6$, $M=0.8$, $M=0.4$, and $M=0.0$, respectively. The color schemes are identical in the top and bottom panels (bottom panel shows less results for clarity). The plasma parameters used in simulations and analytical estimates are: H$^+$ ions, $L_e=20$ and $\tau=10$.}\label{Compar}
\end{figure}

The reduced surface potential of the particle is then obtained by equating the electron and ion fluxes [Eqs. (\ref{Je}) and (\ref{Jeff}), respectively]. As a check of the reliability of the approach developed, we performed calculations for the parameter regime studied by Haakonsen and Hutchinson~\cite{Haakonsen}. The results of this comparison are shown in Fig.~\ref{Compar}. The top panel demonstrates that the proposed approximation provides fairly good agreement with the numerical results in the regimes of weak and strong ion collisionality, but overestimates the ion fluxes in the vicinity of their peaks (except the case of highly supersonic ion drift, where the ion flux is underestimated). The bottom panel shows the corresponding values of the reduced particle charge, calculated from $z=-\tfrac{1}{2}\ln(2\pi m_e j_i^2/m_i)$. The agreement between the numerical simulations and the present analytical approximation is rather good in the entire regime of ion collisionality. We should mention that the numerical results from Ref.~\cite{Haakonsen} are for a particular scenario of ion flow generation. Namely, in simulations ion drift is generated by ion-neutral charge exchange collisions in the presence of drifting neutral background. We expect, however, that the approximation described above is not very sensitive to such details and the essential physics is captured.

The important result, documented in Fig.~\ref{Compar}, is that in the regime of weak and moderate ion collisionality ($\omega_c\lesssim 1$) the ion flux collected by the particle decreases with the ion flow velocity. This decrease is especially pronounced at moderate collisionalities, where significant collisional enhancement of the ion flux is observed in the isotropic case. This enhancement is weakened as the ion flow velocity increases. The physical reason is clear: Higher flow velocities imply higher ion kinetic energies, which makes the region of strong ion-particle interaction narrower, and, as a result, reduces the collisional contribution to the ion flux. Regarding the particle charge, its absolute magnitude increases with the ion flow velocity in the considered regime of weak and moderate ion collisionality (this is of course the consequence of decreasing the flux of positive ions to the particle surface). Note, that the trend will reverse in the regime of strong ion collisionality, which is, however, of no interest for the present study.

The ion-neutral collision frequency $\nu$ has been treated as an independent parameter in this Section. In fact, however, it is generally dependent on the ion drift velocity. To simplify the calculations we will employ the assumption of the constant ion-neutral collision cross section below. The collisional frequency is approximated as
\begin{equation}\label{frequency}
\nu\simeq n_n\sigma \sqrt{v_{T_i}^2+u^2},
\end{equation}
so that $\nu\simeq n_n\sigma v_{T_i}$ in the isotropic regime and $\nu\simeq n_n\sigma u$ in the highly anisotropic regime. The cross section $\sigma\simeq 1\times 10^{-14}$ is chosen, which would reasonably describe the mobility of Ne$^+$ ions in neon in the near-thermal drift regime (see Appendix).

The last important point to mention is that ion-neutral collisions are not the only mechanism leading to the enhanced ion collection. Essentially the same effect can be associated with electron impact ionization events in the vicinity of the particle~\cite{KhrapakPoP2012}. Both mechanisms can be added in a simple superposition and their cumulative effect is $\Delta j_i^{\rm sum}=(1+{\mathcal K})j_i^{\rm coll}$, where ${\mathcal K}=\nu_{\rm I}/\nu$ is just the ratio of ionization and ion-neutral collision frequencies. This ratio is a sharply increasing function of the electron temperature and exceeds unity for $T_e\gtrsim 6.5$ eV in neon plasmas~\cite{KhrapakPoP2012}. Since the electron temperatures up to $T_e\simeq 8$ eV have been measured in the PK-4 facility in the regime of low neutral gas pressures~\cite{KhrapakPRE}, the effect of ionization should not be neglected in the present study. We, therefore, take it into account, when evaluating the particle charge.

\section{Main forces on the particles}\label{Forces}

\begin{figure*}[t!]
\centering
\includegraphics[width=15 cm]{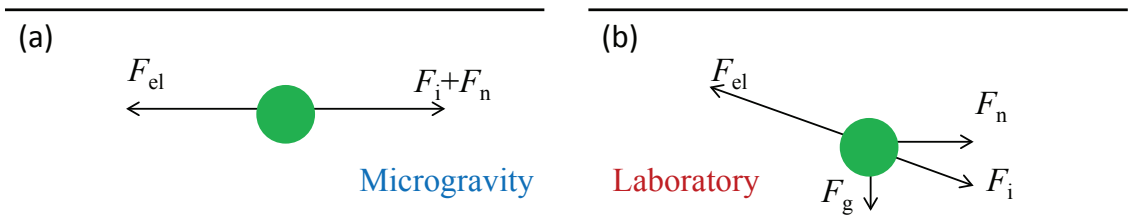}
\caption{(Color online) Sketch of the forces acting on a single particle in microgravity (a) and laboratory (b) conditions. The longitudinal discharge electric field is pointing to the right. Negatively charged particle moves to the left.}\label{Forces_Sketch}
\end{figure*}

The main forces affecting charged grains in complex plasmas subject to an external dc electric field are the electrical force and the forces associated with the momentum transfer from drifting ions and electrons -- ion and electron drag forces. The electrical force is
\begin{equation}\label{Fel}
{\bf F}_{\rm el}= Q{\bf E},
\end{equation}
where the particle charge $Q$ implicitly depends on the electric field via the ion drift velocity $u$. The ion drift velocity and the electric field are related via the ion mobility coefficient, ${\bf u}=\mu {\bf E}$. In the Frost approximation~\cite{Frost} we have
\begin{equation}\label{drift}
u\simeq a(E/p)\left[1+b(E/p)\right]^{-1/2},
\end{equation}
where the coefficients are $a\simeq 1.4\times 10^9$ dyne/statvolt/s and $b\simeq 1.6\times10^4$ dyne/statvolt/cm (in CGS units). The ion flow, caused by the electric field, naturally results in the momentum transfer from the ions drifting relative to the particle component. The resulting force -- the ion drag force~\cite{Kilgore,KhrapakPRE2002,KhrapakPRL2003} -- can be written as
\begin{equation}\label{id}
{\bf F}_{\rm i}=\nu_{id} m_i {\bf u},
\end{equation}
where $\nu_{id}$ is the characteristic momentum transfer frequency in ion-particle collisions. The relative velocity is very close to the actual ion drift velocity since the particles are very massive and their motion can be neglected in comparison with that of the ions. For the momentum transfer frequency we use the result derived in Ref.~\cite{KhrapakPoP2005}
\begin{align}\label{nui}
\begin{split}
\nu_{id} & \simeq \frac{\sqrt{\pi} a^2 n_i v_{T_i}^2}{\xi}\Big{\{}\sqrt{\frac{\pi}{2}}{\rm erf}(\xi) \Big{[}1+2\xi^2+\left(1-\frac{1}{2\xi^2}\right)\times \\ & \times
\left(1+2z\tau\right)+\frac{2z^2\tau^2}{\xi^2}\ln\Lambda \Big{]} +\frac{1}{\sqrt{2}\xi} \Big{[}1+2z\tau+2\xi^2- \\ & -4z^2\tau^2\ln\Lambda \Big{]}e^{-\xi^2} \Big{\}},
\end{split}
\end{align}
where $\ln\Lambda$ is the Coulomb logarithm. This expression is applicable for arbitrary ion drift velocity, but does not take into account the effect of ion-neutral collisions on the ion drag force~\cite{HutchinsonID,Ivlev2004,Ivlev2005,SemenovID}. The latter is relatively small as long as the ion mean free path does not become very short compared to the plasma screening length~\cite{Ivlev2004}. Therefore, for the present purposes we can neglect collisional effects. The Coulomb logarithm reads~\cite{KhrapakPoP2005}
\begin{equation}\label{CL}
\ln\Lambda \simeq \ln \left[ \frac{R_{\rm C}+\lambda}{R_{\rm C}+a}\right],
\end{equation}
where the Coulomb radius $R_{\rm C}$ and the effective plasma screening length $\lambda$ depend on the ion drift velocity as discussed in Sec.~\ref{Model}.

Electric fields can also generate electron drifts (in particular, this occurs in the axial electric field in the positive column of dc discharges, which is relevant to the present study). However, the corresponding electron drag force is usually much smaller than the ion drag force, provided the electron-to-ion temperature ratio is high~\cite{ED}. Thus, the electron drag force is neglected in the present study.

The neutral drag force (Epstein drag~\cite{Epstein}) acts in the direction opposite to the particle motion. It can be written as
\begin{equation}\label{Fn}
{\bf F}_{\rm n}=-\nu_{nd}m_n{\bf V}_d,
\end{equation}
where $m_n$ is the neutral mass, ${\bf V}_d$ is the particle velocity relative to the stationary background of neutrals (gas flows are absent in these experiments), and $\nu_{nd}=(8\sqrt{2\pi}/3)\delta a^2 n_n v_{T_n}$ is the momentum transfer frequency in particle-neutral collisions. The numerical factor $\delta=1+\frac{\pi}{8}\simeq 1.4$ corresponding to diffuse scattering with complete accommodation of neutrals from the particle surface is used, in agreement with the reported experimental results on $\mu$m-size melamine formaldehyde spheres~\cite{Goree}.

Finally, in laboratory experiments particles are naturally affected by the force of gravity, ${\bf F}_{\rm g}= m_d{\bf g}$, where $g\simeq 980$ cm/s$^2$ is the gravitational acceleration.

\section{Force balance in laboratory and microgravity conditions}\label{Force-balance}

The qualitative difference between the balance of forces acting on the particles in laboratory and migrogravity conditions is illustrated in Fig.~\ref{Forces_Sketch}. Under microgravity the particle cloud is located centrally (near the tube axis) and the particles feel only the horizontal electric field $E_{||}$, which also determines the ion drift velocity. In contrast, in laboratory conditions the particle cloud is somewhat shifted downwards from the center (see Fig.~\ref{Flows}). This is obviously because some radial electric field is required, in order for the corresponding electrical force pointing upwards can balance for the particle gravity. At this position the particles feel the {\it total} electric field, which is a sum of the horizontal and radial components, $E=\sqrt{E_{||}^2+E_{\perp}^2}$. The total ion drift velocity at the position of particles is now determined by $E$, not by $E_{||}$ as in microgravity. Since $E>E_{||}$ the ion velocity is effectively higher for particles in ground-based conditions than under microgravity. As has been discussed in Section~\ref{Model}, higher drift velocities imply higher particle charges and therefore higher values of the electrical force. This merely explains why the particle drift velocity should be higher in ground based conditions than in microgravity for the same discharge conditions. (Note that the ratio of the ion drag force to the electrical force remains almost constant for subthermal ion drifts and then decreases rapidly with ion velocity in the suprathermal regime~\cite{KhrapakPoP2005}, and this can be an additional factor contributing to the difference in particle velocities). The discussed effect is clearly more pronounced for bigger particles, since bigger particles imply higher $E_{\perp}$ and hence larger difference between $E$ and $E_{||}$. It also should disappear at sufficiently high pressures, such that ion drifts are subthermal and the particle charge is independent of the ion drift velocity.

Under microgravity conditions, the particles drift velocity can be simply calculated from
\begin{equation}\label{balance_1}
F_{\rm el}^{||} + F_{\rm i}^{||} + F_{\rm n}=0.
\end{equation}
Probe measurements have evidenced that the axial electric field is practically independent of pressure and we therefore take $E_{||} \simeq 2.1$ V/cm as fixed. Then the calculations proceed as follows: (i) The ion drift velocity, as a function of pressure, is calculated from Eq.~(\ref{drift}); (ii)  The ion-neutral collision frequency is calculated from Eq.~(\ref{frequency}); (iii)
The particle charge, as a function of pressure, is evaluated from the model described in Sec.~\ref{Model} (talking into account the dependence of relevant parameters on pressure, ion drift velocity, and collisional frequency); (iv) The ion drag force is calculated from Eqs. (\ref{id})-(\ref{CL}); (v) Solution of Eq.~\ref{balance_1} then yields the dependence of particle drift velocity on pressure.

Regarding laboratory experiments, the calculations are more involved in this case, because the calculation of the particle charge and the ion drag force should now be {\it coupled self-consistently} to the force balance condition in the vertical direction
\begin{equation}\label{balance_2}
F_{\rm el}^{\perp}+F_{\rm i}^{\perp}+F_{\rm g} =0.
\end{equation}
In practice, Equation (\ref{balance_2}) along with the model equations for particle charge and ion drag force are solved to determine $E_{\perp}$ and the total electric field $E$. This fixes the total ion drift velocity, collisional frequency, particle charge, and the ion drag force so that the particle velocity can be evaluated from the force balance condition in the horizontal direction [Eq.~(\ref{balance_1})].

Results from these calculations will be presented in the next Section.

\section{Results}\label{results}

\begin{figure}[t!]
\centering
\includegraphics[width=8 cm]{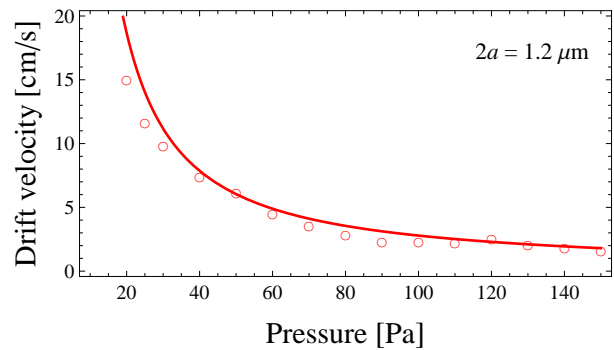}
\caption{(Color online) Particle drift velocity as a function of the neutral gas pressure for particles with diameter $2a=1.2$ $\mu$m. Symbols correspond to the previous (laboratory) experimental results~\cite{RatynskaiaPRL,KhrapakPRE}, solid curve is obtained using the model of the present paper.}\label{1_2mcm}
\end{figure}

We start with presenting results for the case of small particles of diameter $2a\simeq 1.2$ $\mu$m. For such small particles gravity plays almost no role since the radial electric field $E_\perp$ required for the force balance is several times smaller than the axial electric field $E_{||}$ \cite{KhrapakEPL}. For this reason, experiments under microgravity conditions have not been performed for this particle size. Figure~\ref{1_2mcm} shows the comparison between experimental particle velocities measured previously in laboratory~\cite{RatynskaiaPRL,KhrapakPRE} and theoretical calculation using the model presented above (only one curve is shown since the curves for laboratory and microgravity conditions are essentially coinciding). The agreement is convincing, indicating that the necessary ingredients have been properly incorporated into the model.

\begin{figure}[t!]
\centering
\includegraphics[width=8 cm]{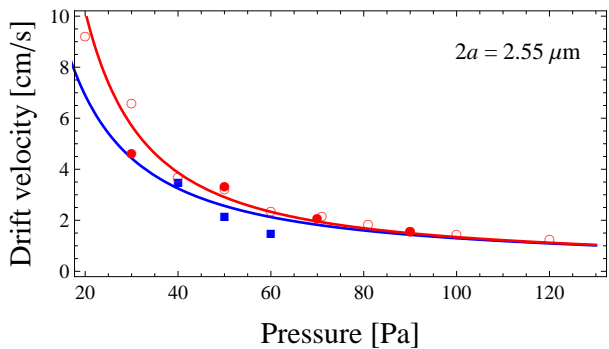}
\caption{(Color online) Particle drift velocity as a function of the neutral gas pressure for particles with diameter $2a=2.55$ $\mu$m. Open (red) circles correspond to the previous laboratory experimental results~\cite{RatynskaiaPRL,KhrapakPRE}. Solid (red) circles correspond to the present laboratory experiment. Solid (blue) squares are the measurements made under microgravity conditions. Upper (red) and lower (blue) curves are obtained using the model of the present paper for laboratory and microgravity conditions, respectively.}\label{2_55mcm}
\end{figure}

Next, we consider larger particles of $2a\simeq 2.55$ $\mu$m in diameter. The results of experiments and calculations are presented in Fig.~\ref{2_55mcm}. The solid (open) circles correspond to the present (previous~\cite{RatynskaiaPRL,KhrapakPRE}) experiments in laboratory. Solid squares are obtained from the analysis of parabolic flights experiments. The upper (red) curve is calculated taking into account gravity. The lower (blue) curve corresponds to calculations for microgravity conditions. Analytical results are in reasonable agreement with the experimental measurements and correctly reproduce the tendency for particles to drift faster in laboratory conditions.

\begin{figure}[t!]
\centering
\includegraphics[width=8 cm]{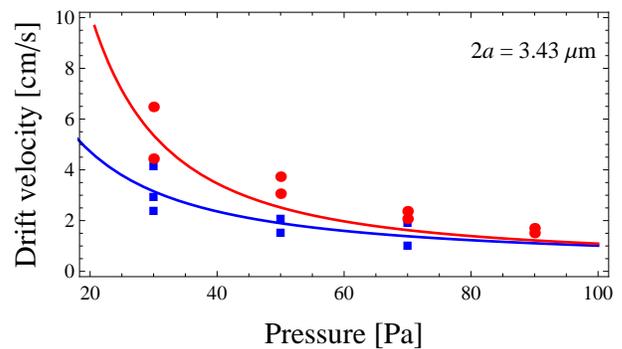}
\caption{(Color online) Particle drift velocity as a function of the neutral gas pressure for particles with diameter $2a=3.43$ $\mu$m. Solid (red) circles correspond to the present laboratory experiment. Solid (blue) squares are the measurements made under microgravity conditions. Upper (red) curve and lower (blue) curves are obtained using the model of the present paper for laboratory and microgravity conditions, respectively.}\label{3_43mcm}
\end{figure}

Finally, we consider the largest particle size investigated, $2a = 3.43$ $\mu$m. For these particles two different dispensers have been used to produce particle injection into the discharge. One is conventional shake dispenser (SD) and the other is the gas-jet dispenser (GJD). It is observed that particles drift faster when GJD is used, the relative difference can amount to $\sim 50\%$~\cite{GJD}.
Nevertheless, clear difference between particle flow velocities in laboratory and microgravity is still observed. This is illustrated in Fig.~\ref{3_43mcm}. Here the solid circles correspond to the laboratory experiments, while solid squares are from microgravity experiments. In each case, upper symbol for a given pressure corresponds to the use of GJD. The upper (red) curve is calculated taking into account gravity. The lower (blue) curve corresponds to calculations for microgravity conditions. We see again that analytical results are in reasonable agreement with the experimental measurements. Comparing Figures \ref{2_55mcm} and \ref{3_43mcm} we also see that the difference in particle drift velocities increases with the particle size and with lowering the neutral gas pressure. Thus the proposed theoretical model adequately describes the main experimental observations.

To conclude this section we comment on the sources of scattering of experimental points in Figs. \ref{1_2mcm} -- \ref{3_43mcm}. The first source is related to uncertainties in measured velocities, which are normally estimated as $10-15 \%$ for the pressure range investigated~\cite{KhrapakPRE,KhrapakEPL}. The second possible source is related to the effect of particles on discharge parameters. The number of injected particles can vary from one experiment to another. Larger numbers of injected particles imply larger modifications of surrounding plasma parameters. Although in the present experiment care has been taken to avoid high particle densities and, therefore, we neglected this effect in the theoretical model, we cannot completely exclude that some variations of plasma parameters do occur. The third source is the systematic difference in particle drift velocities observed when using different dispensers and documented for the largest particle size investigated (see Fig.~\ref{3_43mcm}). This important effect is under current investigation and is presumably associated with long-living neutral flows excited by the gas-jet dispenser~\cite{GJD}.

\section{Conclusion}\label{conclusion}

It has been observed that flows of particles in a positive column of a horizontal dc discharge are characterized by different velocities, depending on whether the experiment is performed in ground-based (laboratory) or microgravity conditions, at otherwise identical discharge parameters. In laboratory particles drift systematically faster than in microgravity and the velocity difference increases when increasing the particle size and/or decreasing neutral gas pressure. In this paper we have provided an explanation for this observation. Qualitatively, gravity shifts the particle downwards from the tube axis, where the radial electric field is strong enough, so that the vertical component of the electrical force can balance for the particle gravity. This region is characterized by faster ion flows and, as a result, the charge of the particles increases compared to the situation where the particles are located close to the tube axis (e.g. in microgravity). Thus, the particles in laboratory experiments feel stronger longitudinal electric force and hence drift faster. To be able to make quantitative comparison between theory and experiments we have developed an analytical model, which properly accounts for the main plasma-particle interaction mechanisms important for the present study. The proposed model yields reasonable agreement with the experimental measurements. It can be of certain value in other situations characterized by relative drifts between the ion and particle components.

\begin{acknowledgements}
We would like to thank C. Haakonsen and I. Hutchinson for providing us with the data from their numerical simulations and reading the manuscript. The PK-4 project is supported by ESA and DLR under
Grant No. 50 WM 1150. We appreciate technical support from C. Deysenroth, C. Rau, and S. Albrecht, and thank ESA for giving us the opportunity to participate
in the 57th ESA parabolic flight campaign. The work was also partly supported by the Russian Foundation for Basic Research, Project No. 13-02-01099.
\end{acknowledgements}

\section*{Appendix: Ion-neutral collision cross section in neon}

We combine the conventional definition of the effective ion-neutral collision frequency $u=eE/\nu m_i$ with Eq.~(\ref{drift}). In the limit of slow drifts ($u\ll v_{T_i}$) we have from Eq.~(\ref{frequency}) $\nu\simeq n_n\sigma v_{T_i}$, which results in $\sigma\simeq ev_{T_i}/a\simeq 1.3\times 10^{-14}$ cm$^2$. In the limit of fast drifts ($u\gg v_{T_i}$) we have $\nu\simeq n_n\sigma u$, which results in $\sigma\simeq e v_{T_i}^2 b/a^2\simeq 0.6\times 10^{-14}$ cm$^2$ (in both cases we assume $T_i\simeq T_n\simeq 0.03$ eV). In this paper we take for the cross section an ``average'' value $\sigma\simeq 1\times 10^{-14}$ cm$^2$, which is a good approximation for near-thermal drifts.


\begin{thebibliography}{99}

\bibitem{Goertz} C. K. Goertz, Rev. Geophys. {\bf 27}, 271 (1989).
\bibitem{Merlino1} R. L. Merlino and J. A. Goree, Phys.Today {\bf 57}, 32 (2004).
\bibitem{Krasheninnikov} S. I. Krasheninnikov, R. D. Smirnov, and D. L. Rudakov, Plasma Phys. Control. Fusion {\bf 53}, 083001 (2011).
\bibitem{Bonitz} M. Bonitz, C. Henning, and D. Block, Rep. Progr. Phys. {\bf 73}, 066501 (2010).
\bibitem{ChaudhuriSM} M. Chaudhuri, A. V. Ivlev, S. A. Khrapak, H. M. Thomas, G. E. Morfill, Soft Matter {\bf 7}, 1287 (2011).
\bibitem{Book1} A. Ivlev, H. L\"{o}wen, G. Morfill, and C. P. Royall, {\it Complex plasmas and colloidal dispersions: Particle-resolved studies of classical liquids and solids} (World Scientific, 2012).


\bibitem{FortovUFN} V. E. Fortov, A. G. Khrapak, S. A. Khrapak, V. I. Molotkov, and O. F. Petrov, Phys. Usp. {\bf 47}, 447 (2004).
\bibitem{FortovPR} V. E. Fortov, A. V. Ivlev, S. A. Khrapak, A. G. Khrapak, and G. E. Morfill, Phys. Rep. {\bf 421}, 1 (2005).
\bibitem{Book2} V. E. Fortov and G. E. Morfill, {\it Complex and dusty plasmas: From Laboratory to Space} (CRC Press, 2010).

\bibitem{KhrapakCPP}  S. Khrapak and G. Morfill, Contrib. Plasma Phys. {\bf 49}, 148 (2009).
\bibitem{MT} S. A. Khrapak, A. V. Ivlev, and G. E. Morfill, Phys. Rev. E {\bf 70}, 056405 (2004).



\bibitem{PK4_a}  M. H. Thoma, M. A. Fink, H. H\"{o}fner, M. Kretschmer, S. A. Khrapak, S. Ratynskaia, V. V. Yaroshenko, G. E. Morfill, O. F. Petrov, A. D. Usachev, A. V. Zobnin, V. E. Fortov, IEEE Transactions Plasma Science {\bf 35}, 255 (2007).
\bibitem{PK4_b}  V. Fortov, G. Morfill, O. Petrov, M. Thoma, A. Usachev, H. H\"{o}fner, A. Zobnin, M. Kretschmer, S. Ratynskaia, M. Fink, K. Tarantik, Yu. Gerasimov and V. Esenkov, Plasma Phys. Controlled Fusion {\bf 47}, B537 (2005).

\bibitem{RatynskaiaPRL}  S. Ratynskaia, S. Khrapak, A. Zobnin, M. H. Thoma, M. Kretschmer, A. Usachev, V. Yaroshenko, R. A. Quinn, G. E. Morfill, O. Petrov and V. Fortov, Phys. Rev. Lett. {\bf 93}, 085001 (2004).
\bibitem{KhrapakPRE}  S. A. Khrapak, S. V. Ratynskaia, A. V. Zobnin, A. D. Usachev, V. V. Yaroshenko, M. H. Thoma, M. Kretschmer, H. H\"{o}fner, G. E. Morfill, O. F. Petrov and V. E. Fortov, Phys. Rev. E {\bf 72}, 016406 (2005).

\bibitem{KhrapakEPL} S. A. Khrapak, P. Tolias, S. Ratynskaia, M. Chaudhuri, A. Zobnin, A. Usachev, C. Rau, M. H. Thoma, O. F. Petrov, V. E. Fortov, and G. E. Morfill, EPL {\bf 97}, 35001 (2012).

\bibitem{YaroshenkoPoP} V. Yaroshenko, S. Ratynskaia, S. Khrapak, M. H. Thoma, M. Kretschmer, H. Hofner, G. E. Morfill, A. Zobnin, A. Usachev, O. Petrov, V. Fortov, Phys. Plasmas {\bf 12}, 093503 (2005).
\bibitem{ZobninPRL}  A. D. Usachev, A. V. Zobnin, O. F. Petrov, V. E. Fortov, B. M. Anaratonne, M. H. Thoma, H. Hofner, M. Kretschmer, M. Fink, G. E. Morfill, Phys. Rev. Lett. {\bf 102}, 045001 (2009).
\bibitem{IvlevPRL2011} A. V. Ivlev, M. H. Thoma, C. Rath, G. Joyce, G. E. Morfill, Phys. Rev. Lett. {\bf 106}, 155001 (2011).
\bibitem{RatynskaiaIEEE} S. Ratynskaia, M. Kretschmer, S. Khrapak, R. A. Quinn, M. H. Thoma, G. E. Morfill, A. Zobnin, A. Usachev, O. Petrov, V. Fortov, IEEE Trans. Plasma Sci. {\bf 32}, 613 (2004).
\bibitem{ZhdanovEPL} S. Zhdanov, R. Heidemann, M. H. Thoma, R. Sutterlin, H. M. Thomas, H. Hofner, K. Tarantik, G. E. Morfill, A. D. Usachev, O. F. Petrov, V. E. Fortov, EPL {\bf 89}, 25001 (2010).
\bibitem{Slobo1} S. Mitic, B. A. Klumov, U. Konopka, M. H. Thoma, and G. E. Morfill, Phys. Rev. Lett. {\bf 101}, 125002 (2008).
\bibitem{Slobo2} S. Mitic, B. A. Klumov, S. A. Khrapak, and G. E. Morfill, Phys. Plasmas {\bf 20}, 043701 (2013).
\bibitem{OverviewPK4} M. H. Thoma, S. Mitic, A. Usachev, B. M. Annaratone, M. A. Fink, V. E. Fortov, H. H\"{o}fner, A. V. Ivlev, B. A. Klumov, U. Konopka, M. Kretschmer, G. E.  Morfill, O. F. Petrov, R. S\"{u}tterlin, S. Zhdanov, A. V. Zobnin, IEEE Trans. Plasma Sci. {\bf 38}, 857 (2010).

\bibitem{Rosenberg} M. Rosenberg, Planet. Space Sci. {\bf 41}, 229 (1993).
\bibitem{FortovPoP_2000} V. E. Fortov, A. G. Khrapak, S. A. Khrapak, V. I. Molotkov, A. P. Nefedov, O. F. Petrov, and V. M. Torchinsky, Phys. Plasmas {\bf 7}, 1374 (2000).
\bibitem{Joyce} G. Joyce, M. Lampe, and G. Ganguli, Phys. Rev. Lett. {\bf 88}, 095006 (2002).
\bibitem{Merlino2} R. Merlino, Phys. Plasmas {\bf 16}, 124501 (2009).

\bibitem{PK4M}   A. Usachev, A. Zobnin, O. Petrov, V. Fortov, M. Thoma, M. Kretschmer, S. Ratynskaia, R. Quinn, H. H\"{o}fner and G. Morfill, Czech. J. Phys. {\bf 54}, C639 (2004).
\bibitem{Zobnin_probes} A. Usachev and A. Zobnin (unpublished).

\bibitem{Zobnin}  A. V. Zobnin, A. P. Nefedov, V. A. Sinelshchikov and V. E. Fortov, JETP {\bf 91}, 483–487 (2000).
\bibitem{Lampe}  M. Lampe, R. Goswami, Z. Sternovsky, S. Robertson, V. Gavrishchaka, G. Ganguli and G. Joyce, Phys. Plasmas {\bf 10}, 1500 (2003).
\bibitem{Hutchinson1}  I. H. Hutchinson and L. Patacchini, Phys. Plasmas {\bf 14}, 013505 (2007).
\bibitem{Dyachkov} L. G. D'yachkov, A. G. Khrapak, S. A. Khrapak, G. E. Morfill, Phys. Plasmas {\bf 14}, 042102 (2007).
\bibitem{Zobnin1} A. V. Zobnin, A. D. Usachev, O. F. Petrov, and V. E. Fortov, Phys. Plasmas {\bf 15}, 043705 (2008).
\bibitem{Interpol} S. A. Khrapak and G. E. Morfill, Phys. Plasmas {\bf 15}, 114503 (2008).
\bibitem{Semenov} I. L. Semenov, A. G. Zagorodny, and I. V. Krivtsun, Phys. Plasmas {\bf 19}, 043703 (2012).

\bibitem{HutchinsonID} L. Patacchini and I. H. Hutchinson, Phys. Rev. Lett. {\bf 101}, 025001 (2008).
\bibitem{Haakonsen} C. B. Haakonsen and I. H. Hutchinson, AIP Conf. Proc. {\bf 1397}, 269 (2011).

\bibitem{OML}   J. E. Allen, Phys. Scr. {\bf 45}, 497 (1992).

\bibitem{KhrapakPoP2012} S. A. Khrapak and G. E. Morfill, Phys. Plasmas {\bf 19}, 024510 (2012).
\bibitem{KhrapakPRE2012} S. A. Khrapak, B. A. Klumov, P. Huber, V. I. Molotkov, A. M. Lipaev, V. N. Naumkin, A. V. Ivlev, H. M. Thomas, M. Schwabe, G. E. Morfill, O. F. Petrov, V. E. Fortov, Yu. Malentschenko, and S. Volkov, Phys. Rev. E {\bf 85}, 066407 (2012).

\bibitem{KhrapakPoP2005} S. A. Khrapak, A. V. Ivlev, S. K. Zhdanov, G. E. Morfill, Phys. Plasmas {\bf 12}, 042308 (2005).
\bibitem{Ludwig} P. Ludwig, W. Miloch, H. K\"{a}hlert, and M. Bonitz, New. J. Phys. {\bf 14}, 053016 (2012).

\bibitem{HutchinsonComment} I. H. Hutchinson, Phys. Plasmas {\bf 14}, 074701 (2007).

\bibitem{Frost} L. S. Frost, Phys. Rev. {\bf 105}, 354 (1957).

\bibitem{Kilgore} M. D. Kilgore, J. E. Daugherty, R. K. Porteous, and D. B.Graves, J.Appl. Phys. {\bf 73}, 7195 (1993).
\bibitem{KhrapakPRE2002}  S. A. Khrapak, A. V. Ivlev, G. E. Morfill and H. M. Thomas, Phys. Rev. E {\bf 66}, 046414 (2002).
\bibitem{KhrapakPRL2003} S. A. Khrapak, A. V. Ivlev, G. E. Morfill, and S. K. Zhdanov, Phys. Rev. Lett. {\bf 90}, 225002 (2003).
\bibitem{Ivlev2004}  A. V. Ivlev, S. A. Khrapak, S. K. Zhdanov, G. E. Morfill, and G. Joyce, Phys. Rev. Lett. {\bf 92}, 205007 (2004).
\bibitem{Ivlev2005}  A. V. Ivlev, S. K. Zhdanov, S. A. Khrapak and G. E. Morfill, Phys. Rev. E {\bf 71}, 016405 (2005).


\bibitem{SemenovID} I. L. Semenov, A. G. Zagorodny, and I. V. Krivtsun, Phys. Plasmas {\bf 20}, 013701 (2013).

\bibitem{ED}   S. A. Khrapak and G. E. Morfill, Phys. Rev. E {\bf 69}, 066411 (2004).

\bibitem{Epstein}   P. S. Epstein, Phys. Rev. {\bf 23}, 710 (1923).

\bibitem{Goree}   Bin Liu, J. Goree, V. Nosenko and L. Boufendi, Phys. Plasmas {\bf 10}, 9 (2003).
\bibitem{GJD} At present we attribute this difference to a gas flow which sets up in the system when using the GJD and survives on time scales needed for the particle transport to the center of the tube.
\end{thebibliography}
\end{document}